\PassOptionsToPackage{numbers,sort&compress}{natbib}

\documentclass{article}

\usepackage[preprint]{neurips_2026}

\usepackage[utf8]{inputenc} 
\usepackage[T1]{fontenc}    
\usepackage{hyperref}       
\usepackage{url}            
\usepackage{booktabs}       
\usepackage{amsfonts}       
\usepackage{nicefrac}       
\usepackage{microtype}      
\usepackage{xcolor}         
\usepackage{amsmath}
\usepackage{graphicx}
\usepackage{array}
\usepackage{multirow}
\usepackage{algorithm}
\usepackage{algpseudocode}
\usepackage{tabularx}

\title{Fairness-Aware and Latency-Controllable Scheduling for Chunked-Prefill LLM Serving}

\author{
  Haoxin Liu \quad Jiayi Wang \quad Yueshen Xu$^*$ \quad Rui Li \\
  \vspace{0.5em} \\
  School of Computer Science and Technology, Xidian University, Xi'an 710071, China \\
  {\small \texttt{\{haoxinliu, jiayiwang\}@stu.xidian.edu.cn}, \quad \texttt{\{ysxu, rli\}@xidian.edu.cn}} \\
  \small{Jiayi Wang and Haoxin Liu contribute equally to this paper, so they are co-first authors.} \\
  \small{*Corresponding author}
}

\begin{document}

\maketitle

\begin{abstract}
	As large language models (LLMs) are increasingly deployed with highly heterogeneous workloads, chunked-prefill execution has emerged as a mainstream serving architecture. Balancing scheduling fairness and latency stability in such environments is critical; otherwise, severe head-of-line blocking and request starvation will degrade user experience. However, existing systems rely on rigid First-Come, First-Served (FCFS) policies and static token budgets, leading to fairness degradation and unpredictable latency jitter. To address these issues, we propose a fairness-aware and latency-controllable scheduling framework for chunked-prefill LLM engines. Specifically, we design a lightweight aging-based scheduling policy that dynamically calculates priorities using accumulated waiting time and remaining prefill work. Furthermore, we develop Latency-Prediction-Based Request Scheduling (LPRS) and Active Prefill Control (APC) to replace static budgets with target-time constraints and actively regulate prefill concurrency. We evaluated our scheduling framework on NVIDIA GPUs and Ascend accelerators using real-world workloads. Results show the aging policy reduces mean end-to-end latency by over 10\% compared to FCFS. Moreover, LPRS and APC significantly reduce P99 tail latency and suppress prefill fragmentation, confirming that the structural prefill control and the temporal latency constraints are fundamentally complementary. All codes have been released in Github\footnote{https://github.com/Charmstok/Chunked\_Prefill\_Serving}\footnote{https://github.com/lhx-666-cool/vllm-ascend/tree/aging}.
\end{abstract}

\section{Introduction}

In recent years, large language models (LLMs) have progressed from proof-of-concept systems to large-scale deployment in applications such as code generation~\cite{zheng2023codegeex}, generative search~\cite{liu2023evaluating}, and retrieval-augmented generation (RAG)~\cite{lewis2020retrieval}. As LLMs are increasingly adopted in real-world serving environments, the performance and scheduling fairness of online inference systems have become critical factors affecting a series of critical issues, including user experience, quality-of-service (QoS), and deployment cost \cite{shi2025deepdiver}. Achieving a balance among low latency, high resource utilization, and fair scheduling has therefore become a central problem in LLM serving. Otherwise, systems may suffer from various adverse effects, including unstable service quality, inefficient resource usage, and even request starvation.

To address these challenges, existing studies have explored several directions: 1) continuous batching and iteration-level scheduling to improve online inference efficiency; 2) chunked prefilling and stall-free batching~\cite{agrawal2024taming} to mitigate prefill-induced blocking on decode; 3) latency-aware scheduling~\cite{sun2024llumnix}, queue management, and fair scheduler based on the continuous batching mechanism~\cite{sheng2024fairness} to better handle request workloads. These efforts have achieved improvements in throughput and latency, and have made chunked-prefill architectures a mainstream design choice. However, most existing methods primarily focus on performance optimization while paying limited attention to scheduling fairness, particularly under mixed workloads. Moreover, the lack of effective coordination between high-level scheduling policies and low-level chunking mechanisms often introduces additional overhead and makes it difficult to balance short-request responsiveness with starvation prevention for long-running jobs. Therefore, designing a scheduling mechanism for chunked-prefill systems that jointly considers multiple critical dimensions such as fairness, efficiency, and practical deployability, remains an insufficiently addressed problem.

In real-world LLM serving scenarios, request streams are rarely homogeneous. Instead, they typically consist of mixed workloads in which short interactive queries, long-context prefilling requests, and long-running generation jobs co-exist within the same batch and contend for shared GPU resources. Crucially, this heterogeneity manifests along two distinct dimensions. The first is intra-batch heterogeneity: within a single scheduling iteration, requests of vastly different lengths and computational profiles must be packed and executed together, forcing the scheduler to reconcile conflicting latency and throughput objectives at fine granularity. The second is inter-batch heterogeneity: the composition of arriving requests fluctuates over time, so that one batch may contain a balanced mixture of short and long requests, while a subsequent batch may be dominated almost entirely by long-context prefilling jobs, or conversely by a burst of short interactive queries. These requests not only differ in latency requirements—short queries demand sub-second responsiveness, whereas long generation jobs tolerate higher end-to-end latency but require sustained throughput—but also exhibit markedly different resource footprints, with prefill-heavy requests being compute-bound and decode-heavy ones being memory-bandwidth-bound. The coexistence of such intra- and inter-batch heterogeneity makes scheduling decisions substantially more challenging than in uniform workload settings, where a one-size-fits-all policy would otherwise suffice.

To bridge this gap, we identify two core under-explored bottlenecks of existing chunked-prefill serving systems: 1) The default First-Come, First-Served (FCFS) scheduling policy can lead to severe fairness degradation, especially in mixed workloads containing both long and short requests, leading to chunk-level head-of-line blocking for short requests and starvation for long requests; 2) The static token budget rule fails to adapt to dynamically changing heterogeneous workloads, resulting in unpredictable latency jitter and low GPU utilization. To address these two pain points, we carry out optimizations from two complementary dimensions: we design a lightweight Aging weighted fair scheduling policy to resolve fairness conflicts, and propose a latency-aware dynamic chunking mechanism with prefill activity control to improve runtime efficiency. Both optimizations are integrated into our scheduling framework, an extended serving system built on top of the open-source Sarathi-Serve~\cite{agrawal2024taming} framework with full compatibility with multi-GPU parallel deployment, and the Aging policy alone is further ported to the Ascend domestic computing platform for industry-specific deployment needs. The main contributions of this paper are summarized as follows:

1. \textbf{We propose a novel Aging weighted fair scheduling policy.} Our policy calculates request priority based on both waiting time and remaining prefill cost, and achieves a smooth trade-off between short request latency optimization and long request starvation prevention with low logarithmic time complexity.

2.\textbf{ We design a new latency-aware adaptive scheduling algorithm}. We design a lightweight latency prediction model that guides a target execution time constraint, replacing the traditional static token budget. We also develop a prefill activity control mechanism to avoid inefficient fragmented execution, effectively reducing latency jitter under heterogeneous workloads.

3. \textbf{We implemented the developed policy and algorithm on two computing architectures \textit{NVIDIA} and \textit{Ascend}}. Our system also supports tensor parallelism and pipeline parallelism in multi-GPU scenarios. Extensive experiments show that our designs  outperform baseline systems in both fairness and throughput performance.

The rest of this paper is organized as follows. Section \ref{Related Work} reviews related work on LLM serving and scheduling. Section \ref{The Developed Methodology} presents the proposed scheduling mechanisms. Section \ref{Experiment} reports the experimental results, and Section \ref{Conclusion and Future Work} concludes the paper and discusses future work.

\section{Related Work}
\label{Related Work}

Modern LLM serving systems have rapidly evolved from batching-centric execution engines to architecture-aware platforms that give the opportunity to optimize scheduling, memory management, or hardware utilization. This section reviews relevant work along three dimensions: 1) LLM serving architectures and chunked-prefill scheduling, 2) cost-model-based latency-aware scheduling, and 3) scheduling and admission control for heterogeneous workloads. 

\subsection{LLM Serving Systems and Chunked-Prefill Scheduling}

LLM serving systems have been widely studied to improve GPU utilization and inference throughput under latency constraints. Early systems mainly rely on continuous batching to increase serving efficiency. Yu et al. \cite{yu2022orca} developed a system Orca, which moves scheduling from the request level to the iteration level and introduces selective batching to mitigate head-of-line blocking. vLLM proposed the paged attention mechanism to improve KV-cache management, reduce memory fragmentation, and support high-throughput serving \cite{kwon2023vllm}.
Recent work further exploits the distinct characteristics of the prefill and decode stages. The prefill stage is compute-intensive and can benefit from large token batches, whereas the decode stage is memory-bandwidth-bound and latency-sensitive. Based on this observation, chunked prefill splits long prefills into smaller schedulable chunks, allowing them to be interleaved with decode operations and reducing the blocking effect of long prompts \cite{agrawal2024taming}. Zhong et al. \cite{zhong2024distserve} proposed a prefill-decode disaggregation approach that places the two stages on separate resources to alleviate cross-stage interference and improve resource utilization. There are some other systems that attempt to further improve serving efficiency for large-scale and long-context inference, through exploring different technical solutions, for example, KV-cache-centric disaggregated architecture, heterogeneous memory management, or hardware-aware optimization  \cite{zhou2024mooncake, shadowkv2025, vllmascend2025, mindie2026llm, zuo2025cloudmatrix, guo2025deepseek}.
Despite these advances, existing systems mainly focus on throughput optimization, memory efficiency, resource partitioning, or hardware adaptation. Request-level fairness under mixed workloads remains less explored, especially in chunked-prefill engines. Although chunking and decode-prioritized execution can reduce cross-stage interference, they may still lead to imbalanced service, excessive waiting time for long requests, or starvation under heterogeneous request patterns. This motivates the requirement for lightweight fairness-aware scheduling policies that can balance throughput, latency, and fairness in LLM serving systems.

\subsection{Cost-model-based latency-aware scheduling}

Prior studies have shown that fixed chunk sizes or static token budgets are often too rigid for online LLM serving, because actual execution latency varies substantially with many reasons, such as batch composition, context length, key-value cache usage, and request concurrency. To address the problem of the high experimental cost associated with optimizing LLM inference deployment under latency SLOs, Agrawal et al. \cite{agrawal2024vidur} presents Vidur, a simulation framework designed for LLM inference performance. It combines experimental profiling with predictive modeling to estimate execution latencies, enabling system operators to identify cost-effective deployment setups without relying on extensive physical trials. These methods allow schedulers to estimate the latency impact of different batching and chunking decisions more accurately, thereby improving throughput and latency stability under dynamic workloads, alongside broader system efficiency and cost-effectiveness.
In an industrial survey, Liu et al. \cite{LiuACMSurvey2026} further validate that production online LLM services generally face challenges such as bursty request arrivals and highly dynamic workload compositions, thereby amplifying the demand for accurate latency prediction and flexible, SLO-aware scheduling.
Despite these benefits, existing cost-centric approaches predominantly optimize system efficiency or average SLO satisfaction. They usually treat cost models as capacity-control tools, leaving the coordination between fairness-oriented policies and the underlying chunked-prefill mechanism largely unexplored.

\subsection{Scheduling and Admission Control for Heterogeneous Workloads}

To address the heterogeneity of real-world LLM workloads, workload analyses from platforms like LMSYS by Zheng et al. \cite{zheng2024distributed} highlight that online services commonly process requests with highly divergent prompt and generation lengths alongside complex multi-tenant demands. In response, existing literature has generally split into two orthogonal directions. At the high level, scheduling systems such as those proposed by Sheng et al. \cite{sheng2024fairness} explore queue-aware policies and admission control to mitigate head-of-line blocking and ensure tenant fairness. At the low level, parallel execution frameworks proposed by researchers from Peking University and Shanghai AI Lab, namely LoongServe \cite{wu2024loongserve}, design fine-grained mechanisms like elastic sequence parallelism to adapt to iteration-level resource demands across prefill and decode phases. While both directions have individually improved serving efficiency, a significant gap remains between high-level scheduling and low-level execution. High-level fairness and priority mechanisms \cite{sheng2024fairness} typically operate at a coarse granularity, restricted to the queue or instance level, thereby failing to capture the fine-grained execution dynamics of the underlying engine. Conversely, low-level parallel serving systems \cite{wu2024loongserve} lack SLO-aware or multi-tenant priority coordination. Without bridging the gap between high-level priority assignment and low-level chunked execution, heterogeneous workloads in chunked-prefill systems remain vulnerable to several critical vulnerabilities, namely excessive memory fragmentation, high scheduling overhead, and severe starvation risks.

While the above studies provide foundations for online LLM inference, they predominantly focus on optimizing throughput, latency, SLO satisfaction, hardware adaptation, or cluster-level deployment efficiency. In contrast, this work addresses a vital but under-explored problem: how to design a lightweight fairness-aware scheduling mechanism for mixed workloads.

\section{The Developed Methodology}
\label{The Developed Methodology}

\subsection{Aging-Based Scheduling for Chunked Prefill}

\subsubsection{Motivation and Design Goals}

Chunked-prefill execution can improve LLM serving by breaking a long prefill into multiple schedulable chunks, allowing prefill and decode requests to be interleaved at a finer granularity. However, this execution model also exposes a new fairness problem. That is, since decode requests are typically scheduled first to preserve generation continuity, only the residual token budget in each round is available to prefill requests. Under this setting, the scheduling order of prefill chunks directly determines how different requests share service over time.

A simple FCFS policy is poorly suited to this fine-grained execution model and the reasons are as follows. On the one hand, once a long prompt reaches the head of the queue, it may repeatedly consume the residual prefill budget across multiple rounds. Short prompts arriving behind it can therefore experience long TTFT (Time to First Token) even though their prefill cost is small. On the other hand, a shortest-prefill-first policy improves the responsiveness of short requests, but it can indefinitely delay long requests when short requests keep arriving. These two policies represent opposite extremes: FCFS preserves arrival order but can amplify chunk-level head-of-line blocking, while shortest-prefill-first reduces short-request latency but weakens starvation protection.

Aging is designed to provide a lightweight middle ground. Its goal is to favor short prefills and to further improve responsiveness, while ensuring that long-waiting requests continuously gain scheduling priority. This policy follows three design principles: First, it must preserve the decode-first scheduling discipline and therefore should not interfere with TPOT (Time Per Output Token) stability. Second, it should be non-intrusive, modifying only the ordering of prefill candidates rather than the chunking mechanism, KV-cache management, or execution backend. Third, it should expose a continuous tradeoff between latency and fairness, instead of relying on hard queue partitioning or static resource reservation.

\subsubsection{Priority Formulation}

For each prefill request $i$, let $a_i$ denote its arrival time, $r_i(n)$ denote the remaining number of prefill tokens at the scheduling round $n$, and $P_i(n)$ denote its scheduling priority. The aging policy assigns priority as
\begin{equation}
	P_i(n) = \alpha \cdot (t - a_i) + \beta \cdot r_i(n),
    \label{eq:priority}
\end{equation}
where $t$ is the current system time, and $\alpha > 0$ is the aging weight, and $\beta < 0$ is the remaining-work weight. The first term $\alpha(t-a_i)$ increases with the waiting time of a request. It gives older requests progressively higher priority and prevents them from being permanently dominated by newly arrived short requests. The second term $\beta r_i(n)$ favors requests with less remaining prefill work, since a smaller $r_i(n)$ incurs a smaller negative penalty. This term preserves the latency benefit of prioritizing short prefills, which is important for interactive workloads.

The two terms together encode the desired fairness behavior. A short request can be served early because its remaining-work penalty is small. A long request may initially have lower priority, but its waiting-time term keeps increasing as it remains in the queue. Therefore, the scheduler does not rely on a fixed request class or a hard starvation threshold; instead, priority evolves continuously with both waiting time and prefill progress.
After request $i$ receives a prefill chunk of size $c_i$, its remaining prefill tokens are updated as
\begin{equation}
	r_i(n+1) = r_i(n) - c_i.
\end{equation}
This update is important for chunked-prefill serving. A long request is not treated as a monolithic job with a fixed cost. As its chunks are gradually processed, its remaining-work penalty decreases, making the request increasingly likely to finish in subsequent rounds. The aging policy therefore couples fairness with chunk-level progress rather than applying a static priority throughout the request lifetime.

\subsubsection{Latency-Fairness Tradeoff and Scheduling Semantics}

\textbf{Latency-fairness tradeoff}. The behavior of the aging policy is controlled by the ratio $\alpha / |\beta|$. 1) When this ratio is small, the remaining-work term dominates the priority function. The scheduler  behaves closer to the shortest-prefill-first policy, so it can give priority to short requests, and thus can reduce TTFT for latency-sensitive workloads. 2) When $\alpha / |\beta|$ is large, the waiting-time becomes more influential. So long requests can accumulate priority more quickly and eventually overtake newly arrived short requests, thus reducing starvation risk under sustained mixed arrivals. Inspired by the above observations, an expectation can be inferred that intermediate values could provide a smooth tradeoff between short-request responsiveness and long-request progress.

The free value setting of $\alpha / |\beta|$ is useful in online LLM serving considering that the length of the requests contained in the workload can vary significantly across applications. Interactive requests may prefer stronger short-request responsiveness, while multi-tenant or batch-heavy deployments may require stronger starvation protection. The aging policy supports all cases through parameter tuning without changing the underlying execution engine.

\textbf{Scheduling semantics.} Aging is applied only after decode requests have been admitted. In each scheduling round, the scheduler first reserves capacity for ongoing decode requests to preserve generation continuity. It then ranks prefill candidates according to the aging priority and allocates the remaining prefill budget in that order.
For a selected request $i$, the scheduled chunk size is bounded by its remaining prefill length and the available prefill budget in the current round. Once the chunk is scheduled, the request state is updated. If the request still has remaining prefill tokens, it is returned to the prefill queue with an updated priority; otherwise, it leaves the prefill stage.

This scheduling semantics makes the aging policy non-intrusive. The policy does not change how chunks are executed, how KV cache is managed, or how batches are submitted to the backend. It only changes which prefill request receives the next scheduling opportunity. As a result, Aging can be integrated into existing chunked-prefill engines as a scheduling-layer replacement for FCFS.

\subsubsection{Efficient Priority Maintenance}

If we rely solely on Eq.~\ref{eq:priority} to implement priority-based scheduling, A naive implementation would recompute $P_i(n)$ for every waiting request in every scheduling round. This is unnecessary because the current time $t$ is shared by all waiting requests within the same round. The priority can be rewritten as
\begin{equation}
	P_i(n) = \alpha \cdot t + \left(-\alpha \cdot a_i + \beta \cdot r_i(n)\right).
\end{equation}
Since $\alpha \cdot t$ is shared by all requests, it does not affect their order. Therefore, the scheduler only needs to maintain the equivalent ordering key:
\begin{equation}
	K_i(n) = -\alpha \cdot a_i + \beta \cdot r_i(n).
\end{equation}

The waiting prefill queue is implemented as a max-heap ordered by $K_i(n)$. When a new request arrives, its key is computed from its arrival time and initial prefill length, and the request is inserted into the heap. When an unfinished request receives a chunk, only its remaining token count and heap key need to be updated before reinsertion.

If the queue contains $n$ requests, and $k$ requests are selected in a scheduling round, the heap operations introduce $O(k \log n)$ overhead. Since $k$ is bounded by the number of prefill requests actually scheduled in that round, the developed aging policy avoids full-queue priority recomputation and keeps scheduling overhead low. Overall, Aging turns prefill scheduling from a one-dimensional arrival-order rule into a configurable fairness-aware policy, while preserving the execution semantics and efficiency of chunked-prefill serving.

\subsection{Latency-Prediction-Based Request Scheduling}
\label{LPRS}

In chunked-prefill LLM serving systems, the scheduler should determine how Prefill and Decode requests are combined in each dynamic batching round. Existing systems commonly rely on a fixed chunk size or a fixed token budget to bound the per-round batch size. Although such designs are simple, they are fundamentally static approximations. They treat token count as a universal proxy for execution cost, while ignoring that the actual batch latency is usually jointly affected by many factors such as decode concurrency, context-length distribution, historical prefill progress, and GPU memory state~\cite{agrawal2024vidur}. Consequently, the same token budget may correspond to substantially different execution time under real online workloads.

This mismatch becomes particularly problematic under mixed Prefill-Decode workloads. On the one hand, when the batch is underfilled, GPU resources will be underutilized; on the other hand, when the batch is overfilled, the execution time of the current round may exceed the scheduling window, causing token-level latency jitter and degraded QoS~\cite{patel2024splitwise}. Therefore, the key problem is not simply how many more tokens can be packed into the current round. Instead, the problem is how to choose a chunk such that the actual execution time remains close to a target latency budget.

Based on this observation, we propose the \textbf{Latency-prediction-based request scheduling (LPRS)} mechanism. The key idea is to shift the scheduling objective from maximizing token filling to minimizing deviation from a target latency budget. Specifically, we first train a lightweight latency predictor offline to estimate the execution time of a candidate batch from its runtime state. During online scheduling, the scheduler performs a discrete search over candidate chunk sizes and selects the one whose predicted latency is closest to the target budget.

\subsubsection{The Developed Offline Latency Predictor}

To provide reliable timing estimates for online scheduling, we first build a batch-latency predictor for a single scheduling round, which consists of four components.

\textbf{1. Raw feature representation.} The developed predictor takes a 16-dimensional feature vector as input, consisting of 11 raw statistics and 5 derived features. These 11 raw features are grouped into three primary categories (computation load, context and memory access, and GPU memory state) and are detailed in Table~\ref{tab:raw_features}.

\begin{table}[!htb]
\centering
\footnotesize
\caption{Summary of 11 raw input features.}
\label{tab:raw_features}
\begin{tabularx}{\columnwidth}{l X}
\toprule
\multicolumn{1}{c}{\textbf{Feature Name}} & \multicolumn{1}{c}{\textbf{Definition \& Role}} \\
\midrule
\multicolumn{2}{l}{\textbf{Computation Load}} \\
\quad \texttt{prefill\_tokens} & Total prefill tokens to be processed. \\
\quad \texttt{decode\_tokens} & Total decode tokens in the batch. \\
\quad \texttt{batch\_request\_count} & Active batched requests in the current round. \\
\midrule
\multicolumn{2}{l}{\textbf{Context \& Memory Access}} \\
\quad \texttt{sum\_decode\_context\_len} & Cumulative context length of decode requests. \\
\quad \texttt{max\_decode\_context\_len} & Maximum context length among decode requests. \\
\quad \texttt{prefill\_processed\_tokens} & Total tokens processed in prefill phase. \\
\quad \texttt{max\_prefill\_processed\_tokens} & Max prefill tokens processed. \\
\midrule
\multicolumn{2}{l}{\textbf{GPU Memory State}} \\
\quad \texttt{gpu\_mem\_used\_mb} & Physical GPU memory utilized (MB). \\
\quad \texttt{gpu\_mem\_free\_mb} & Physical GPU memory currently free (MB). \\
\quad \texttt{cuda\_allocated\_mb} & Memory allocated by CUDA memory pool (MB). \\
\quad \texttt{cuda\_reserved\_mb} & Total memory reserved by CUDA pool (MB). \\
\bottomrule
\end{tabularx}
\end{table}

\textbf{2. Derived feature construction.} In addition to these raw statistics, we construct five derived features to better capture the non-linear coupling between the Prefill-Decode phase structure and the actual execution cost. Their definitions and roles are detailed in Table~\ref{tab:LPRS_derived_features}.

\begin{table}[!htb]
\centering
\footnotesize
\caption{Derived features used by the latency predictor.}
\label{tab:LPRS_derived_features}
\begin{tabular}{
  >{\centering\arraybackslash}p{0.30\linewidth}
  >{\raggedright\arraybackslash}p{0.65\linewidth} 
}
\toprule
\multicolumn{1}{c}{\textbf{Feature Name}} & \multicolumn{1}{c}{\textbf{Definition \& Role}} \\ 
\midrule
\multirow{2}{*}[-2.5ex]{\texttt{bias}} & \multicolumn{1}{c}{$1$} \\
& Captures fixed overhead even when dynamic features are close to zero, such as kernel launch overhead, scheduling framework overhead, and GPU/CPU/NPU synchronization cost. \\ \midrule
\multirow{2}{*}[-0.4ex]{\texttt{scheduled\_tokens}} & \multicolumn{1}{c}{$\texttt{decode\_tokens} + \texttt{prefill\_tokens}$} \\
& Represents the overall workload intensity of the current batch. \\ \midrule
\multirow{2}{*}[-0.4ex]{\texttt{avg\_decode\_ctx}} & \multicolumn{1}{c}{$\displaystyle \frac{\texttt{sum\_decode\_context\_len}}{\max(\texttt{decode\_tokens}, 1)}$} \\
& Reflects the average context depth of the current decode requests. \\ \midrule
\multirow{2}{*}[-1ex]{\texttt{decode\_ctx\_interaction}} & \multicolumn{1}{c}{$\texttt{decode\_tokens} \times \texttt{avg\_decode\_ctx}$} \\
& Captures the interaction between decode-token generation volume and average context length. \\ \midrule
\multirow{2}{*}[-1ex]{\texttt{prefill\_interaction}} & \multicolumn{1}{c}{$\texttt{prefill\_tokens} \times \texttt{prefill\_processed\_tokens}$} \\
& Jointly characterizes the scale of current prefill tokens and the depth of their historical context. \\
\bottomrule
\end{tabular}
\end{table}

\textbf{3. Data collection and preprocessing.} We collect offline training data by running the token-budget scheduler on prompts under three measures including diverse arrival rates, prompt-length mixtures, and concurrency levels. For each scheduling round, the system records the 11 raw features together with the ground-truth batch latency measured by a high-resolution timer. The raw samples are cleaned, and bucketed by total scheduled tokens. They are then downsampled for overrepresented full-chunk cases, and split into training, validation, and test sets.

\textbf{4. Model architecture and training.} The latency predictor is designed as a lightweight three-layer multilayer perceptron (MLP) with hidden dimensions, rectified linear unit (ReLU) activations, and a dropout rate. The model is trained using a weighted asymmetric Huber loss:
\begin{equation}
L=\frac{1}{N}\sum_{i=1}^{N} w_i\, l_i(y_i, \hat{y}_i),
\end{equation}
where $w_i$ denotes the bucket-aware sample weight, and $l_i(y_i, \hat{y}_i)$ represents the asymmetric Huber loss for the $i$-th sample given its ground-truth latency $y_i$ and predicted latency $\hat{y}_i$. This loss could improve robustness to outliers through the Huber base while imposing a stronger penalty on latency underestimation~\cite{dabney2018distributional}. It is expected to reduce the risk of budget overflow during online scheduling.

\subsubsection{Optimization, Candidate Search, and Scoring}

\textbf{Optimization.} Consider a candidate prefill request $i$ in scheduling round $t$, let $r_i$ denote its remaining prompt tokens, $B_{\max}$ denote the hard token budget per round, and $U_t$ denote the total tokens already committed to the current batch. The feasible upper bound of the chunk size is derived as:
\begin{equation}
h_i=\min(r_i,\; B_{\max}-U_t).
\end{equation}

In the original static scheduler, the chunking decision for request $i$ may directly adopts the feasible upper bound $h_i$ in the current round as the final chunk allocation $c_i^{\text{base}}$:
\begin{equation}
\label{eq:original_scheduler}
c_i^{\text{base}}=h_i.
\end{equation}

This strategy is equivalent to greedily maximizing token filling. However, it ignores the fact that identical chunk sizes may induce substantially different execution time under varying context states. To address this issue, the developed LPRS preserves the hard token constraint while introducing a target latency budget $T^\ast$, thereby reformulating the scheduling process as a time-deviation minimization problem.

\textbf{Candidate Search and Scoring.} To reduce online search overhead, we employ a discrete search instead of exhaustive enumeration. Let $\Delta$ denote the search granularity. Assuming $h_i \ge 1$ (otherwise the request is skipped with $c_i^\ast=0$), the candidate set is constructed as:
\begin{equation}
\mathcal{C}_i=\{1,h_i\}\cup\{k \cdot \Delta \mid 1\le k \cdot \Delta\le h_i\}.
\end{equation}

For each candidate $c\in\mathcal{C}_i$, the scheduler constructs the corresponding 16-dimensional feature vector $x_{t,i}(c)$ and queries the predictor $f_\theta(\cdot)$ to estimate the batch execution time:
\begin{equation}
\hat{T}_{t,i}(c)=f_\theta(x_{t,i}(c)).
\end{equation}

We measure the fitness of each candidate with respect to the target latency budget using the following asymmetric scoring function:
\begin{equation}
\text{Score}_{t,i}(c)=
\begin{cases}
\lambda_u\left(T^\ast-\hat{T}_{t,i}(c)\right), & \hat{T}_{t,i}(c)\le T^\ast, \\[4pt]
\lambda_o\left(\hat{T}_{t,i}(c)-T^\ast\right), & \hat{T}_{t,i}(c)>T^\ast.
\end{cases}
\end{equation}

Here, $\lambda_u$ and $\lambda_o$ represent the penalties for underfilling and overflow, respectively. We set $\lambda_o>\lambda_u$ to reflect that exceeding the latency budget is strictly more harmful to system stability than leaving a small margin of computation capacity unused. The final chunk decision is thus:
\begin{equation}
c_i^\ast=\arg\min_{c\in\mathcal{C}_i}\text{Score}_{t,i}(c).
\end{equation}

The online procedure of LPRS consists of four steps. First, the system initializes the round state and collects GPU memory statistics. Second, it prioritizes running decode requests to preserve decode continuity. Third, it performs a discrete candidate search for active prefills and newly arrived requests in the waiting queue, scoring each candidate according to its predicted latency. Finally, the system finalizes the batch structure and submits it to the backend for execution. The key innovation of LPRS is that it no longer optimizes solely for token filling. Instead, it explicitly aims to keep the batch execution time close to the target latency budget. To achieve this, LPRS introduces a time-budget-aware chunk selection mechanism, as detailed in Algorithm \ref{alg:select_prefill_chunk}.

\begin{algorithm}[htbp]
\caption{The Proposed Algorithm for Selecting Prefill Chunk by Time Budget}
\label{alg:select_prefill_chunk}
\small
\begin{algorithmic}[1]
\Require
    Candidate prefill request $i$ with $r_i$ remaining prompt tokens, total committed tokens $U_t$, hard token budget $B_{\max}$, search granularity $\Delta$, target latency budget $T^\ast$
\Ensure
    Selected chunk decision $c_i^\ast$
\State Compute upper bound:
    \Statex \hspace{1.5em} $h_i \gets \min(r_i,\; B_{\max} - U_t)$
\If{$h_i \leq 0$}
    \State \Return 0
\EndIf
\State Construct candidate set:
    \Statex \hspace{1.5em} $\mathcal{C}_i \gets \{1, h_i\} \cup \{k\Delta \mid 1 \leq k\Delta \leq h_i\}$
\State $\text{best\_score} \gets +\infty$
\State $c_i^\ast \gets 0$
\For{each $c \in \text{sorted}(\mathcal{C}_i)$}
    \State Construct feature vector $x_{t,i}(c)$ assuming chunk $c$ is selected
    \State Estimate batch latency: $\hat{T}_{t,i}(c) \gets f_\theta(x_{t,i}(c))$
    \If{$\hat{T}_{t,i}(c) \leq T^\ast$}
        \State $\text{Score}_{t,i}(c) \gets \lambda_u \left(T^\ast - \hat{T}_{t,i}(c)\right)$
    \Else
        \State $\text{Score}_{t,i}(c) \gets \lambda_o \left(\hat{T}_{t,i}(c) - T^\ast\right)$
    \EndIf
    \If{$\text{Score}_{t,i}(c) < \text{best\_score}$}
        \State $\text{best\_score} \gets \text{Score}_{t,i}(c)$
        \State $c_i^\ast \gets c$
    \ElsIf{$\text{Score}_{t,i}(c) == \text{best\_score}$ and $c > c_i^\ast$}
        \State $c_i^\ast \gets c$
    \EndIf
\EndFor
\State \textit{// Prevent starvation for empty batch}
\If{$c_i^\ast == 0$ and $U_t == 0$ and $h_i \geq 1$}
    \State \Return 1
\EndIf
\State \Return $c_i^\ast$
\end{algorithmic}
\end{algorithm}

\subsection{The Designed Active Prefill Control Mechanism}

Although LPRS controls the time budget of each scheduling round, it does not directly regulate the structure of unfinished prefills. Under decode-dominated high-contention workloads, the system may keep too many unfinished prefills active simultaneously. \textbf{As a result, the limited token budget may be fragmented into many tiny chunks with little practical progress.} Conversely, if the system becomes overly decode-biased, new requests may suffer from prolonged starvation. To address this issue, we introduce an \textbf{Active Prefill Control (APC)} mechanism on top of target-latency scheduling to regulate unfinished-prefill activity and progress granularity.

\subsubsection{Dynamic Activity Cap}

Active prefill concurrency control must regulate the maximum number of active, incomplete prefill requests in a batch. Under heavy mixed workloads, uncontrolled concurrency leads to two primary inefficiencies: 1) \textbf{budget dilution}, where the remaining token budget ($B_{\max} - U_t$) is divided among too many requests, resulting in highly fragmented, sub-optimal chunk sizes; and 2) \textbf{micro-progress}, where requests are allocated negligible chunk sizes (e.g., a single token), trivially maintaining their active status while inflating the total number of scheduling rounds. In both cases, the minimal forward progress is heavily offset by fixed scheduling overheads, such as feature extraction, latency prediction, and kernel launch latency.

Consequently, concurrency control must go beyond binary admission decisions to dynamically determine the exact number of active prefill requests. Because decoding loads and available memory slots fluctuate across scheduling rounds, a static limit is insufficient. We therefore introduce a dynamic concurrency upper bound that adapts to the real-time batch state, throttling prefill concurrency during heavy decoding phases and increasing active requests during idle periods to maximize hardware utilization.

Let $\mathcal{D}_t$ denote the decode set and $\mathcal{P}_t$ denote the unfinished-prefill set in round $t$. The maximum number of active prefills is derived as:
\begin{equation}
\label{eq:DynamicActivityCap}
C_t=\min\left(
C_{\max},
S_{\max}-|\mathcal{D}_t|,
\left\lfloor\frac{B_{\max}-U_t}{L_{\min}}\right\rfloor
\right).
\end{equation}

Here, $C_{\max}$ is the configured maximum number of active prefills, and $S_{\max}$ is the maximum sequence capacity. $B_{\max}$ is the hard token budget, $U_t$ is the number of already committed tokens, and $L_{\min}$ is the minimum effective chunk size for each active prefill. This formulation ensures that the active-prefill concurrency is jointly constrained by sequence slots, the remaining token budget, and the minimum progress quality.

\subsubsection{Minimum Effective Progress and Warm Start}

Having established the dynamic concurrency upper bound, we must also constrain the minimum chunk size for individual prefill requests. This boundary prevents scenarios where request concurrency remains compliant, but the actual forward progress of each request is too minor to be effective. For any unfinished prefill request $i$, the minimum effective progress is derived as:
\begin{equation}
m_i=\min(r_i, L_{\min}),
\end{equation}
where $r_i$ is the remaining prompt length and $L_{min}$ is identical to the same symbol in Eq. \ref{eq:DynamicActivityCap}. If the chunk $c_i^\ast$ proposed by LPRS satisfies both the activity-cap constraint and the minimum-progress constraint, APC will directly accept it. Otherwise, if the current batch contains no active prefill, APC will trigger a warm-start step to restore nonzero prefill activity at minimal cost. The decision rule is
\begin{equation}
c_i=
\begin{cases}
c_i^\ast, & \text{if } |\mathcal{P}_t|<C_t \text{ and } c_i^\ast\ge m_i, \\[4pt]
\min(h_i,m_i), & \text{if } c_i^\ast<m_i \text{ and } |\mathcal{P}_t|=0, \\[4pt]
0, & \text{otherwise.}
\end{cases}
\end{equation}

Intuitively, APC favors keeping a small number of meaningful prefills active, rather than allowing a large number of low-quality fragmented chunks to occupy the active set for a long time.

\begin{figure}[htbp]
\centering
\includegraphics[width=\linewidth]{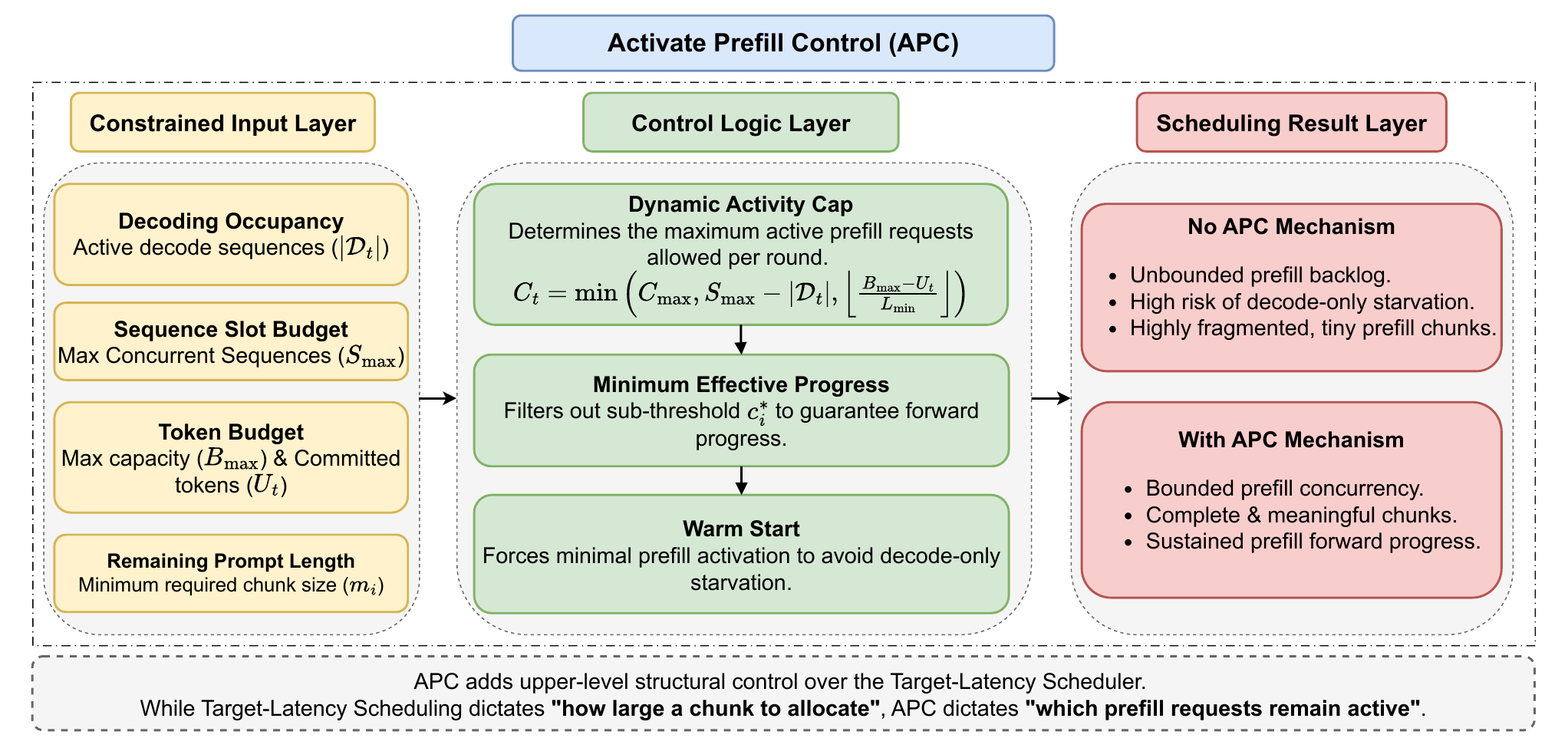}
\caption{The workflow of the designed Active Prefill Control mechanism.}
\label{fig:apc_mechanism}
\end{figure}

Figure~\ref{fig:apc_mechanism} summarizes the control logic of the designed APC. It first determines the admissible number of active unfinished prefills according to the current decode occupancy and remaining token budget (yellow boxes). It then checks whether the chunk proposed by LPRS satisfies the minimum effective progress requirement (green boxes). When no active prefill exists in the current batch, APC will invoke the warm-start rule to restore nonzero prefill activity with minimal scheduling cost (green boxes). Through this control structure, the system preserves a small number of viable prefill forward progress opportunities even when decoding dominates, thereby mitigating fragmented advances and maintaining a non-zero prefill activity (red boxes). 

\section{Experiment and Evaluation}
\label{Experiment}

\subsection{Experimental Setup}

All experiments are implemented on top of LLM serving systems with chunked-prefill execution. We evaluate three sets of scheduling mechanisms: The aging policy is evaluated as an independent fairness-oriented prefill scheduling policy. LPRS and APC are evaluated separately for latency-budget control and active-prefill regulation. Unless otherwise stated, experiments use Qwen3-8B\footnote{https://huggingface.co/Qwen/Qwen3-8B} as the test model.

For the aging policy, the single-GPU experiments were implemented based on Sarathi-Serve and run on one NVIDIA RTX 4090 GPU. The maximum context length is set to 512, and each request has a maximum generation length of 512 tokens. The workload is replayed from ShareGPT\footnote{https://huggingface.co/datasets/anon8231489123/ShareGPT\_Vicuna\_unfiltered} with 100 or 200 requests and a fixed 0.1s request inter-arrival interval. ShareGPT is a dataset of real-world multi-turn conversations collected from interactions with large language models, and is widely used for evaluating LLM serving systems under realistic workloads. This workload exhibits a highly skewed prompt-length distribution: for the 200-request setting, the median prompt length is 19.0 tokens while the P90 prompt length reaches 179.4 tokens, making it suitable for evaluating scheduling fairness under mixed short- and long-prompt requests.

For LPRS and APC, we employed Qwen3-8B, and set both the tensor parallelism and pipeline parallelism to 1. That is, the two parallelisms were not used. The hardware platform is a single NVIDIA RTX PRO 5000 GPU with 72GB VRAM, running Ubuntu 24.04.4 LTS. The latency-prediction profiling dataset contains 36{,}868 samples, which are split into training, validation, and test sets at an 8:1:1 ratio. We tested LPRS on two online workloads: a high-concurrency scenario with 0.1s request inter-arrival interval, and a regular scenario with 1.0s interval, each containing 1{,}000 requests with maximum sequence length 4{,}096 tokens. For APC ablation, we used a heterogeneous workload with a 49:1 short-prompt (30-50 tokens) to long-prompt (200-220 tokens) ratio. The arrival of requests did not follow a fixed rate, but could change dynamically. The statistics for the data and workloads for LPRS and APC are presented in Table~\ref{table:Dataset-and-workload-statistics}. All codes are released in Github in two repositories\footnote{https://github.com/Charmstok/Chunked\_Prefill\_Serving}\footnote{https://github.com/lhx-666-cool/vllm-ascend/tree/aging}. 

\begin{table}[htbp]
\centering
\small
\caption{Dataset and workload statistics.}
\label{table:Dataset-and-workload-statistics}
\begin{tabular}{@{}lr p{1cm} lr@{}}
\toprule
\textbf{Parameter / Item} & \textbf{Value} && \textbf{Parameter / Item} & \textbf{Value} \\
\midrule
Offline profiling samples      & 36{,}868        && LPRS regular interval          & 1.0 s           \\
Train/Val/Test split           & 8:1:1           && APC short prompt length        & 30--50 tokens   \\
Test samples                   & 3{,}713         && APC long prompt length         & 200--220 tokens \\
Online requests per workload   & 1{,}000         && APC short:long ratio           & 49:1            \\
LPRS high-concurrency interval & 0.1 s           &&                                &                 \\
\bottomrule
\end{tabular}
\end{table}
\subsection{Evaluation Metrics}

\textbf{Latency prediction evaluation}. We evaluate the latency predictor using Mean Absolute Error (MAE), Root Mean Squared Error (RMSE), and Mean Absolute Percentage Error (MAPE):
\begin{equation}
\text{MAE}=\frac{1}{N}\sum_{i=1}^{N} |\hat{y}_i-y_i|,
\end{equation}
\begin{equation}
\text{RMSE}=\sqrt{\frac{1}{N}\sum_{i=1}^{N}(\hat{y}_i-y_i)^2},
\end{equation}
\begin{equation}
\text{MAPE}=\frac{100\%}{N}\sum_{i=1}^{N}\left|\frac{\hat{y}_i-y_i}{y_i}\right|.
\end{equation}
Here, $y_i$ and $\hat{y}_i$ denote the ground-truth latency and the predicted latency, respectively. We also report percentile absolute errors to assess tail robustness, because tail prediction quality is particularly important for latency-sensitive online scheduling.

\textbf{End-to-end serving performance evaluation}. We report request-level and prefill-level latency. For request $q$, the full end-to-end (E2E) latency and the prefill E2E latency are defined as
\begin{equation}
L_q^{\text{req}} = t_q^{\text{finish}} - t_q^{\text{arrive}},
\end{equation}
\begin{equation}
L_q^{\text{pf}} = t_q^{\text{prefill\_done}} - t_q^{\text{arrive}}.
\end{equation}
We summarize these metrics with percentile statistics, because high-percentile latency is more informative than mean latency in interactive LLM serving.

\subsection{Evaluation for the Aging Policy}

The developed aging policy is evaluated as an independent fairness-oriented prefill scheduling policy. Its goal is to reduce queueing delay under mixed short- and long-prompt workloads while preserving the execution semantics of chunked-prefill serving. We compare the aging policy with the default FCFS policy, which orders prefill requests only by arrival time.

\subsubsection{Fairness under Mixed Workloads}

We first evaluate the aging policy under the 200-request mixed workload. The workload is dominated by short prompts but contains a small number of substantially longer requests. This distribution stresses the scheduler because long requests can repeatedly consume residual prefill budget across multiple scheduling rounds, which will cause short requests behind them to wait even when their prefill cost is small.

\begin{table}[htbp]
  \centering
  \small
  \setlength{\tabcolsep}{3pt}
  \caption{Aging vs. FCFS under the 200-request mixed workload.}
  \label{tab:aging-single-gpu}
  \begin{tabular}{crrrrr}
    \hline
    Chunk & Policy & Mean E2E & P95 E2E & Mean TTFT & P95 TTFT \\
    \hline
    256  & FCFS  & 118.72s & 195.73s & 107.82s & 184.86s \\
    256  & \textbf{Aging} & \textbf{106.56s} & \textbf{183.58s} & \textbf{95.67s}  & \textbf{172.73s} \\
    512  & FCFS  & 115.52s & 196.63s & 104.12s & 185.24s \\
    512  & \textbf{Aging} & \textbf{107.73s} & \textbf{185.72s} & \textbf{96.72s}  & \textbf{174.78s} \\
    1024 & FCFS  & 109.38s & 190.10s & 98.14s  & 178.34s \\
    1024 & \textbf{Aging} & \textbf{110.59s} & \textbf{189.64s} & \textbf{99.39s}  & \textbf{178.39s} \\
    \hline
  \end{tabular}
\end{table}

\begin{figure}[htbp]
  \centering
  \includegraphics[width=0.95\linewidth]{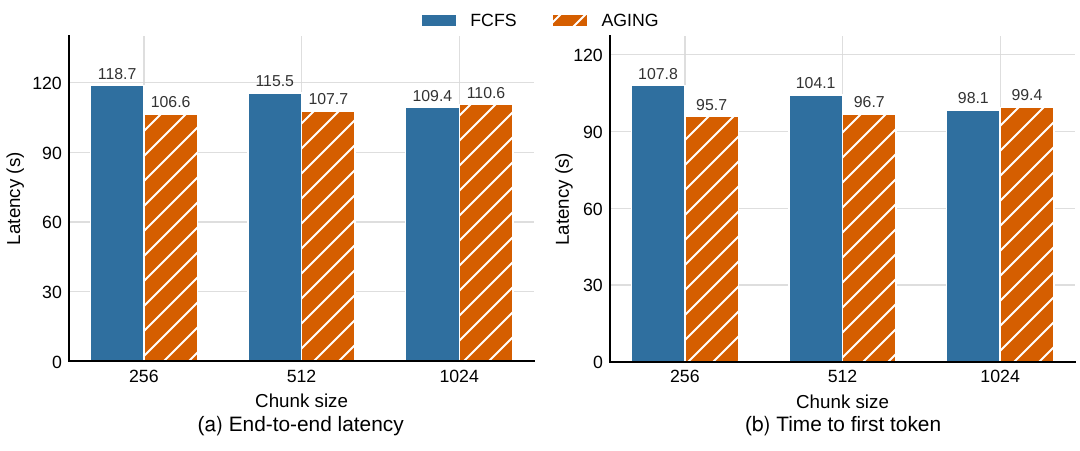}
  \caption{Mean E2E latency and TTFT under FCFS and Aging.}
  \label{fig:policy-latency}
\end{figure}

\begin{figure}[htbp]
  \centering
  \includegraphics[width=0.95\linewidth]{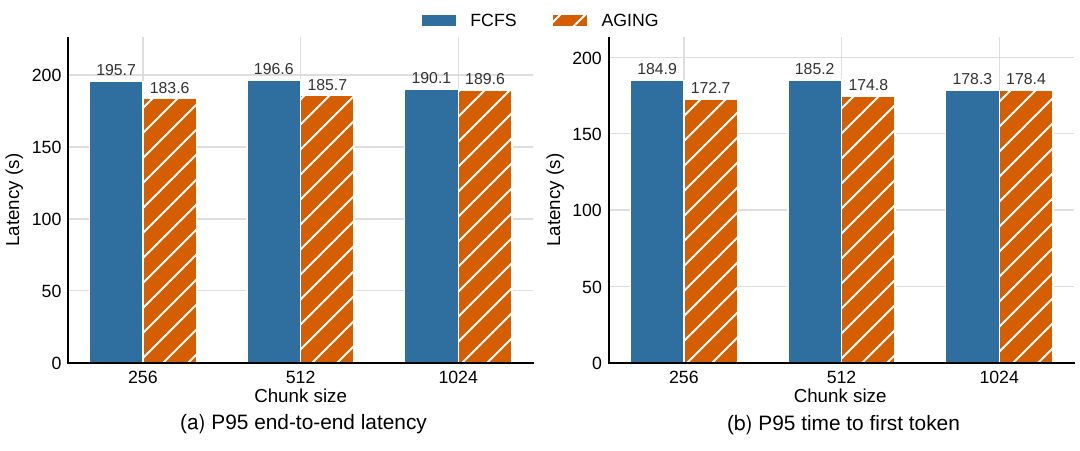}
  \caption{P95 E2E latency and TTFT under FCFS and Aging.}
  \label{fig:tail-latency}
\end{figure}

Table~\ref{tab:aging-single-gpu} reports the single-GPU results under different chunk sizes. With a 256-token chunk, our aging policy reduces mean request end-to-end latency from 118.72s to 106.56s, achieving 10.24\% improvement over FCFS. Mean TTFT decreases from 107.82s to 95.67s, improving by 11.27\%. Tail latency also improves: E2E P95 decreases from 195.73s to 183.58s, and TTFT P95 decreases from 184.86s to 172.73s. With a 512-token chunk, our aging policy still improves mean E2E latency by 6.75\% and improves mean TTFT by 7.11\%. Figs.~\ref{fig:policy-latency} and~\ref{fig:tail-latency} visualize the same trend and we can draw the following conclusions.

\begin{enumerate}
    \item Aging consistently shifts both mean latency and tail latency downward when the chunk size is 256 or 512 tokens. This indicates that the improvement is not limited to average-case requests, that is, the long-waiting tail also benefits from the fairness-aware ordering.
    \item The benefit becomes smaller when the chunk size increases to 1,024 tokens. In this case, Aging and FCFS show nearly identical P95 latency, while Aging slightly increases mean E2E latency. This behavior is expected: Aging relies on the scheduler that has enough opportunities to reorder prefill requests. Larger chunks will reduce scheduling granularity and limit the ability of the priority rule to intervene once a long chunk has been admitted.
    \item To understand where the improvement comes from, we decompose latency under the 256-token chunk setting. Aging does not reduce model execution time: the average model execution time is 10.85s under Aging and 10.86s under FCFS. Instead, the improvement comes almost entirely from queuing reduction. The average scheduling waiting time decreases from 107.77s to 95.63s. This confirms that Aging improves latency by changing prefill service order rather than accelerating model execution.
\end{enumerate}

\begin{figure}[htbp]
  \centering
  \includegraphics[width=0.72\linewidth]{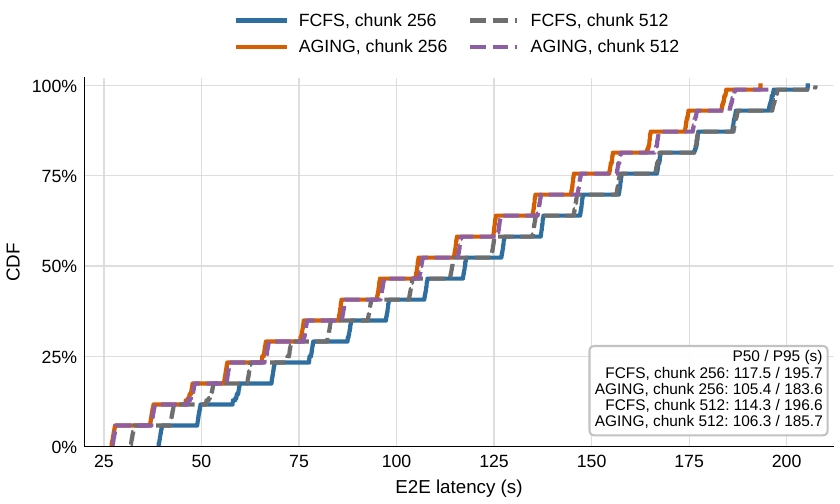}
  \caption{CDF of request E2E latency under FCFS and Aging.}
  \label{fig:e2e-cdf}
\end{figure}

Fig.~\ref{fig:e2e-cdf} further shows the E2E latency distribution. The Cumulative Distribution Function (CDF) of the aging policy is shifted to the left of FCFS for most requests, indicating that the scheduling policy improves latency for the majority of the workload rather than only optimizing a small subset of requests.

\subsubsection{Parameter Sensitivity}

\begin{figure}[htbp]
  \centering
  \includegraphics[width=0.72\linewidth]{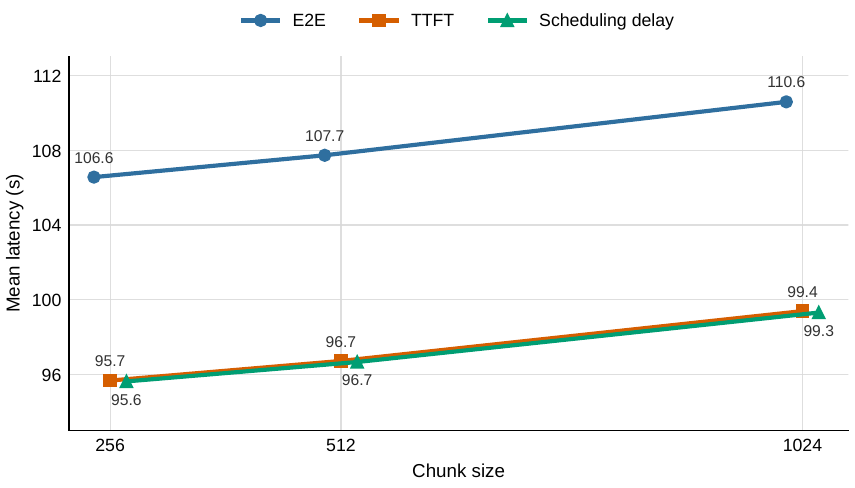}
  \caption{Sensitivity of the aging policy to the chunk size.}
  \label{fig:chunk-sensitivity}
\end{figure}

We study the sensitivity of the aging policy to two key parameters: chunk size and the waiting-time weight. Chunk size controls the granularity at which the scheduler can revise prefill order. Fig.~\ref{fig:chunk-sensitivity} reports the experimental results. Under the 200-request workload, the mean E2E latency of Aging increases from 106.56s to 107.73s and 110.59s as the chunk size grows from 256 to 512 and 1024 tokens. Mean TTFT follows the same trend, increasing from 95.67s to 96.72s and 99.39s.  Smaller chunks expose more scheduling opportunities, allowing Aging to react more quickly when short requests wait behind long prompts.

\begin{figure}[htbp]
  \centering
  \includegraphics[width=0.72\linewidth]{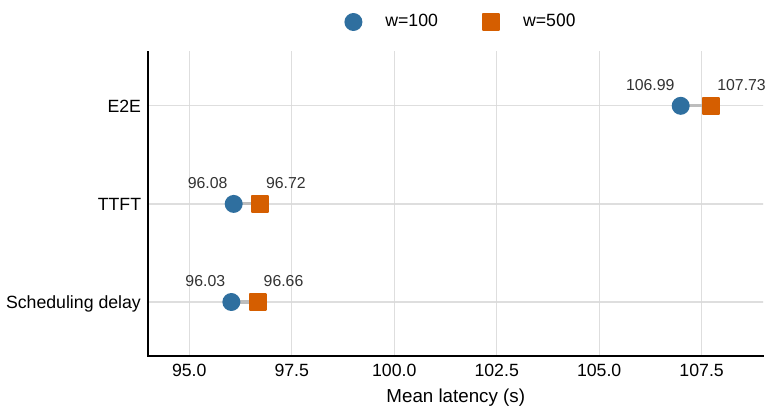}
  \caption{Sensitivity of Aging to the waiting-time weight.}
  \label{fig:weight-sensitivity}
\end{figure}

The waiting-time weight controls how aggressively old requests accumulate priority. Using the chunk size of 512, we compare two waiting-weight settings. When the weight base is 100, mean E2E latency is 106.99s and TTFT P95 is 173.20s. Increasing the weight base to 500 raises mean E2E latency to 107.73s and TTFT P95 to 174.78s. Fig.~\ref{fig:weight-sensitivity} shows that a larger waiting-time weight does not improve latency in this workload because it weakens the influence of the remaining-work term, making the policy closer to pure time-based ordering. This result highlights that Aging should balance short-request preference with starvation prevention; overly aggressive aging can reduce the responsiveness benefit for short prompts.

\subsubsection{Multi-GPU Extension}

We further evaluate whether Aging remains compatible with multi-GPU execution. The experiment uses two NVIDIA RTX 5090 GPUs and the same 200-request ShareGPT workload. Aging reuses the centralized scheduling design of Sarathi-Serve: all priority computation and request ordering are performed at the engine side, and the resulting schedule is broadcast to worker GPUs through the existing communication path. This design avoids distributed priority queues and preserves the global schedule consistency required by tensor and pipeline parallelism.

Table~\ref{tab:aging-multigpu} shows the two-GPU results. With the chunk size of 256 and the sequence capacity of 10, Aging performs worse than FCFS because the limited concurrency prevents its waiting-time compensation from translating into useful scheduling progress. Increasing the sequence capacity to 32 substantially reduces latency for both policies. Under the best tested configuration, the chunk size of 512 and the sequence capacity of 32, Aging reduces mean E2E latency from 61.77s to 60.55s and mean TTFT from 46.16s to 45.34s. The improvement is modest, but the result confirms that Aging can run correctly under multi-GPU execution without modifying the underlying parallel runtime.

\begin{table}[htbp]
  \centering
  \small
  \caption{Performance of the aging policy under two-GPU execution.}
  \label{tab:aging-multigpu}
  \begin{tabular}{@{}cclrrrr@{}}
    \toprule
    \multicolumn{2}{c}{\textbf{Configuration}} & \multirow{2}{*}{\textbf{Policy}} & \multicolumn{2}{c}{\textbf{E2E Latency (s)}} & \multicolumn{2}{c}{\textbf{TTFT (s)}} \\
    \cmidrule(lr){1-2} \cmidrule(lr){4-5} \cmidrule(lr){6-7}
    \textbf{Chunk Size} & \textbf{Max Seqs} & & \textbf{Mean} & \textbf{P95} & \textbf{Mean} & \textbf{P95} \\

    \midrule
    \multirow{2}{*}{256} & \multirow{2}{*}{10}
    & FCFS  & 99.05 & 165.72 & 89.47 & 156.27 \\
    & & Aging & 103.84 & 173.58 & 93.79 & 163.83 \\
    \midrule
    
    \multirow{2}{*}{256} & \multirow{2}{*}{32}
    & FCFS  & 63.89 & 93.45 & 48.61 & 83.49 \\
    & & Aging & 64.20 & 95.96 & 48.29 & 84.07 \\
    \midrule
    
    \multirow{2}{*}{512} & \multirow{2}{*}{32}
    & FCFS  & 61.77 & 93.13 & 46.16 & 81.52 \\
    & & Aging & 60.55 & 92.03 & 45.34 & 79.80 \\
    \bottomrule
  \end{tabular}
\end{table}

\subsubsection{Portability in the Huawei Ascend Ecosystem}

Finally, we study whether our aging policy can be deployed on a Huawei accelerator platform. We port the policy to vLLM-Ascend\footnote{https://github.com/vllm-project/vllm-ascend} on a single Ascend 910B
NPU\footnote{https://e.huawei.com/cn/products/computing/ascend} and run an OpenPangu-family model\footnote{https://gitcode.com/ascend-tribe}  \cite{chen2025panguembeddedefficientdualsystem} with 200 ShareGPT requests. The port keeps the core Aging logic unchanged. Platform-specific changes are limited to the scheduling integration layer, including configuration injection, policy registration, and chunk-size alignment required by the Ascend backend. The code has been released in Github.

Table~\ref{tab:aging-ascend} reports the measured performance. Aging completes the 200-request workload in 14.99s, achieving 13.35 requests/s throughput. The E2E latency distribution is stable, with P95 latency of 10.52s and P99 latency of 12.70s. Since this experiment is intended as a portability validation rather than a cross-hardware performance comparison, we do not compare absolute latency against the NVIDIA results. The result shows that Aging can be deployed on a non-NVIDIA serving stack through lightweight scheduling-layer integration.

\begin{table}[htbp]
\centering
\small
\setlength{\tabcolsep}{7pt} 
\caption{Performance of the aging policy on Ascend 910B with vLLM-Ascend.}
\label{tab:aging-ascend}
\begin{tabular}{@{}cc|ccccccc@{}} 
\toprule
\multicolumn{2}{c}{\textbf{Global Metrics}} & \multicolumn{7}{c}{\textbf{End-to-End Latency Profile (s)}} \\
\cmidrule(lr){1-2} \cmidrule(lr){3-9}
\textbf{Total Time (s)} & \textbf{Throughput (req/s)} & \textbf{Mean} & \textbf{P50} & \textbf{P80} & \textbf{P90} & \textbf{P95} & \textbf{P99} & \textbf{Max} \\
\midrule
14.99 & 13.35 & 6.546 & 6.523 & 8.599 & 9.560 & 10.520 & 12.700 & 12.900 \\
\bottomrule
\end{tabular}
\end{table}

\subsection{Evaluation for the Designed LPRS}

This section evaluates the target-latency scheduling mechanism from two dimensions: predictor accuracy and scheduling performance. 1) Predictor accuracy primarily assesses whether the latency prediction model can leverage the decoding/prefill statistical features of the current batch, context lengths, and GPU runtime states, to provide reliable estimates of candidate batch execution latencies. 2) Regarding scheduling performance, we focus on whether the scheduler, which is aided by latency predictions, can allocate appropriate chunks to pending prefill requests under hard token constraints. This is designed to align the actual execution time of each iteration more closely with the target budget, rather than simply saturating the remaining token capacity based on Eq. \ref{eq:original_scheduler} as is employed in the baseline scheduler. 

\subsubsection{Accuracy Evaluation for Latency Predictor}

We first detail the specific model configuration and optimization parameters of the lightweight MLP predictor. The model is trained using the AdamW optimizer \cite{loshchilov2017decoupled} under the hyperparameter settings summarized in Table~\ref{tab:mlp_hyperparameters}.

\begin{table}[htbp]
\centering
\small
\caption{Hyperparameter configurations of the MLP latency predictor.}
\label{tab:mlp_hyperparameters}
\begin{tabularx}{\columnwidth}{l l X}
\toprule
\textbf{Hyperparameter} & \textbf{Default Value} & \textbf{Description} \\
\midrule
\texttt{epochs} & 300 & Total number of training epochs. \\
\texttt{lr} & $2 \times 10^{-3}$ & Learning rate of the AdamW optimizer. \\
\texttt{weight\_decay} & $1 \times 10^{-3}$ & Weight decay coefficient for AdamW regularization. \\
\texttt{hidden\_sizes} & $(128, 64, 32)$ & Dimensionality of the hidden layers. \\
\texttt{dropout} & 0.1 & Dropout probability. \\
\texttt{batch\_size} & 256 & Mini-batch size. \\
\bottomrule
\end{tabularx}
\end{table}

Following the training process, we evaluate the predictor's generalization capability on the held-out test set. Table~\ref{tab:predictor_accuracy} reports the prediction accuracy of the lightweight MLP predictor on the test set. The model achieves an MAE of 1.13 ms, an RMSE of 1.53 ms, and an MAPE of 1.26\%. Notably, 99.00\% of the predictions exhibit an absolute error of no more than 5 ms, with all samples bounded within 10 ms. These results demonstrate that the predictor delivers the high precision required for stable online scheduling.

\begin{table}[htbp]
\centering
\small
\setlength{\tabcolsep}{8pt}
\caption{Accuracy of the latency predictor on the test set.}
\label{tab:predictor_accuracy}
\begin{tabular}{@{}ccccccccc@{}}
\toprule
\multicolumn{3}{c}{\textbf{Overall Error}} & \multicolumn{4}{c}{\textbf{Absolute Error Percentiles (ms)}} & \multicolumn{2}{c}{\textbf{Bounded Error Ratio}} \\
\cmidrule(lr){1-3} \cmidrule(lr){4-7} \cmidrule(lr){8-9}
\textbf{MAE} & \textbf{RMSE} & \textbf{MAPE} & \textbf{P50} & \textbf{P90} & \textbf{P95} & \textbf{P99} & \textbf{$\boldsymbol{\le 5}$ ms} & \textbf{$\boldsymbol{\le 10}$ ms} \\
\midrule
1.13 ms & 1.53 ms & 1.26\% & 0.84 & 2.49 & 3.18 & 4.93 & 99.00\% & 100.00\% \\
\bottomrule
\end{tabular}
\end{table}

\subsubsection{End-to-End Performance Evaluation}

To evaluate the efficacy of target-latency control, we compare LPRS (Section~\ref{LPRS}) against a fixed token-budget baseline. This comparison aims to determine whether our approach improves latency stability when static token budgets fail to accurately reflect actual execution costs under heavy mixed workloads. We conduct evaluations under two distinct traffic patterns: \textit{regular load}, characterized by uniform request arrivals at a 1.0~s interval, and \textit{high concurrency}, where requests arrive uniformly at a compressed 0.1~s interval. Table~\ref{tab:LPRS_latency_comparison} reports the prefill and full-request E2E latencies under these conditions.  

\begin{enumerate}
   \item Under high concurrency, LPRS successfully keeps most mixed Decode-Prefill batches close to the 105~ms target budget, yielding consistently superior tail latency compared to the baseline. Conversely, under regular load, the performance gap narrows, as the static token policy already operates near the system's natural efficiency point.
   \item Specifically, under high concurrency (0.1~s interval), LPRS reduces the P99 prefill latency from 924.22~ms to 889.45~ms, and the P99 request latency from 986.93~ms to 952.56~ms. Under regular load (1.0~s interval), however, both methods deliver similar latency profiles. Notably, the prefill tail latency (P99) under regular load exhibits a sharp jump to over 22~ms compared to the sub-millisecond P90 baseline. This tail is primarily attributed to the non-preemptive nature of iteration-level scheduling: a rare prefill request arriving mid-iteration must wait for the active decode round to finish, thereby incurring a scheduling quantization delay.
\end{enumerate}

\begin{table}[htbp]
\centering
\small
\caption{E2E latency comparison between LPRS (105~ms target) and Token Budget (1{,}024) scheduling.}
\label{tab:LPRS_latency_comparison}
\setlength{\tabcolsep}{12pt}
\begin{tabular}{l rr cc}
\toprule
\multirow{2}{*}{\textbf{Quantile}} & \multicolumn{2}{c}{\textbf{Prefill Latency (ms)}} & \multicolumn{2}{c}{\textbf{Full Request Latency (ms)}} \\
\cmidrule(lr){2-3} \cmidrule(lr){4-5}
 & \textbf{LPRS} & \textbf{Token Budget} & \textbf{LPRS} & \textbf{Token Budget} \\
\midrule
\multicolumn{5}{l}{\textit{High Concurrency (0.1~s arrival interval)}} \\
P50 & 397.88 & 406.09 & 545.99 & 555.51 \\
P80 & 647.75 & 662.35 & 797.26 & 830.74 \\
P90 & 772.10 & 814.88 & 908.34 & 942.81 \\
P99 & 889.45 & 924.22 & 952.56 & 986.93 \\
\midrule
\multicolumn{5}{l}{\textit{Regular Load (1.0~s arrival interval)}} \\
P50 & 0.0934 & 0.0890 & 138.52 & 129.29 \\
P80 & 0.1208 & 0.1174 & 150.60 & 147.47 \\
P90 & 0.1446 & 0.1353 & 152.49 & 147.83 \\
P99 & 22.890 & 22.010 & 153.35 & 148.99 \\
\bottomrule
\end{tabular}
\end{table}

\subsection{Evaluation for the Designed APC}

To evaluate the contribution of APC, we compare LPRS without the APC mechanism (APC off) and the normal LPRS (APC on) under the same scheduling framework. Table~\ref{tab:apc_ablation} presents the results. When APC is enabled, the average number of scheduled prefill sequences decreases dramatically from 5.32 to 0.46, whereas the average prefill chunk size surges from 0.78 to 6.29. This behavior clearly indicates that APC reduces fragmentation by suppressing ineffective concurrency, rather than by blindly increasing prefill parallelism.

At the latency level, APC reduces the mean request E2E latency by 22.26\% and the P90 request E2E latency by 24.62\%. Similarly, mean and P90 prefill E2E latencies drop by 23.04\% and 25.01\%, respectively. As expected, enabling APC introduces non-zero intervention events (4,960 blockings by the activity cap and 1,541 by the minimum-effective-chunk rule). Far from degrading performance, these calculated interventions are aligned with APC's design objective: it prevents low-value prefill fragments from occupying the active set, thereby significantly improving overall scheduling efficiency. These results verify that structural prefill control is complementary to temporal target-latency control.

\begin{table}[htbp]
\centering
\small
\caption{Ablation results of the APC mechanism.}
\label{tab:apc_ablation}
\begin{tabular}{@{}lrrr@{}}
\toprule
\textbf{Metric} & \textbf{APC Off} & \textbf{APC On} & \textbf{Change} \\
\midrule
\multicolumn{4}{@{}l}{\textit{End-to-End Latency}} \\
Mean Request E2E (ms) & 143.65 & 111.68 & -22.26$\%$ \\
Mean Prefill E2E (ms) & 136.04 & 104.69 & -23.04$\%$ \\
P90 Request E2E (ms) & 254.89 & 192.15 & -24.62$\%$ \\
P90 Prefill E2E (ms) & 247.58 & 185.66 & -25.01$\%$ \\
\midrule
\multicolumn{4}{@{}l}{\textit{Scheduling Dynamics}} \\
Avg. Scheduled Prefill Seqs. & 5.32 & 0.46 & -91.37$\%$ \\
Avg. Prefill Chunk Size & 0.78 & 6.29 & +706.96$\%$ \\
\midrule
\multicolumn{4}{@{}l}{\textit{Intervention Events (Counts)}} \\
Blocked by Activity Cap & 0 & 4{,}960 & -- \\
Blocked by Min. Effective Chunk & 0 & 1{,}541 & -- \\
\bottomrule
\end{tabular}
\end{table}

\subsection{Discussion}
The experiments evaluate three scheduling mechanisms under the broad setting of chunked-prefill LLM serving, and we can draw the following conclusions:

1) Aging focuses on fairness-oriented prefill ordering. Its benefit appears when mixed workloads contain many short prompts and a small number of long prompts: by combining remaining prefill work with accumulated waiting time, Aging reduces queuing delay and improves both mean and tail latency when the scheduling granularity is sufficiently fine. The sensitivity results also reveal its boundary. When chunks become too large, the scheduler has fewer opportunities to reorder prefill requests, and the benefit may diminish.

2) LPRS addresses a different problem. latency-budget control under dynamic mixed workloads. Its benefit is most visible under high contention, where a fixed token budget no longer reliably reflects actual execution time. By selecting chunks according to predicted latency, LPRS improves batch execution-time controllability and reduces tail latency. 

3) APC further studies the structure of active prefills. By limiting ineffective prefill concurrency and increasing useful chunk progress, APC reduces fragmentation and improves request latency.

Overall, these results suggest that chunked-prefill serving requires scheduling policies that are aware of multiple sources of inefficiency. Arrival-order scheduling alone is insufficient for fairness under mixed prompt lengths; static token budgets are insufficient for latency control under dynamic batch composition; and unconstrained active-prefill concurrency can create fragmentation. The experiments therefore demonstrate that improving LLM serving requires reasoning not only about batching efficiency, but also about fairness, latency controllability, and prefill structure.

\section{Conclusion and Future Work}
\label{Conclusion and Future Work}

This paper investigates fairness-aware and latency-controllable scheduling for chunked-prefill LLM serving. We proposed three lightweight scheduling mechanisms: an aging-based scheduler for fairness-aware request prioritization, LPRS for latency-prediction-guided chunk selection, and APC for regulating active unfinished prefills. These mechanisms address the limitations of arrival-order scheduling, static token-budget allocation, and unconstrained active-prefill concurrency, respectively, especially under mixed online workloads. These mechanisms improve the chunked-prefill scheduling process from the perspectives of request ordering, execution-time control, and prefill progress structure, while requiring only limited modifications to existing serving systems. Overall, this work shows that efficient chunked-prefill LLM serving should not rely solely on static batching policies, but should jointly consider request heterogeneity, runtime latency variation, and scheduling progress quality, to improve fairness, latency stability, and hardware utilization. We also performed sufficient experiments under varying computing architectures (Nvidia and Ascend), and the results verify the effectiveness and efficiency of the developed mechanisms under all test enviroments and cases. 

Future work can be extended in two directions. First, a unified adaptive scheduling controller can be developed to jointly coordinate Aging, LPRS, and APC, and to dynamically adjust scheduling parameters according to changing online workloads. Second, the proposed mechanisms should be further evaluated in larger-scale and more complex production environments, including larger models, longer-context workloads, multi-node deployments, or multi-tenant settings, in order to further examine their scalability and practical applicability.

\bibliographystyle{IEEEtran}
\bibliography{ref}
\end{document}